\documentclass[aps,prb,reprint,superscriptaddress,showpacs,showkeys,amsmath,floatfix,longbibliography]{revtex4-1}

\usepackage{color}
\usepackage{graphicx}% Include figure files
\usepackage{dcolumn}% Align table columns on decimal point
\usepackage{bm}% bold math
\usepackage{textcomp}
\usepackage{verbatim}
\usepackage{hyperref}% add hypertext capabilities
\begin{document}

\title{Direct Observation of Attractive Skyrmions and Skyrmion Clusters in the Cubic Helimagnet Cu$_2$OSeO$_3$ }

\author{J.~C.~Loudon}
\email{j.c.loudon@gmail.com}
\affiliation{Department of Materials Science and
  Metallurgy, 27 Charles Babbage Road, Cambridge, CB3 0FS, United Kingdom.}
	
\author{A.~O.~Leonov}
\email{leonov@hiroshima-u.ac.jp}
\affiliation{Department of Chemistry, Faculty of Science, Hiroshima University
Kagamiyama, Higashi-Hiroshima, Hiroshima 739-8526, Japan.}
\affiliation{Chiral Research Center, Hiroshima University, Higashi-Hiroshima, 739-8526, Japan.}
\affiliation{IFW Dresden, Postfach 270016, D-01171 Dresden, Germany.}   

\author{A.~N.~Bogdanov}
\affiliation{Chiral Research Center, Hiroshima University, Higashi-Hiroshima, 739-8526, Japan.}
\affiliation{IFW Dresden, Postfach 270016, D-01171 Dresden, Germany.} 

\author{M. Ciomaga Hatnean}
\author{G. Balakrishnan}
\affiliation{Department of Physics, University of Warwick, Coventry CV4 7AL, United Kingdom.}

\date{\today}

%%%%%%%%%%%%%%%%%%%%%%%%%%%%%%%%%%%%%%%%%%%%%\

\begin{abstract}

We report the discovery of \textit{attractive} magnetic skyrmions and their clusters in non-centrosymmetric ferromagnets. These three-dimensional solitons have been predicted to exist in the cone phase of chiral ferromagnets (J. Phys: Condens. Matter \textbf{28} (2016) 35LT01) and are fundamentally different from the more common repulsive axisymmetric skyrmions that occur in the magnetically saturated  state. We present real-space images of these skyrmion clusters in thin ($\sim 70$~nm) single-crystal samples of Cu$_2$OSeO$_3$ taken using transmission electron microscopy and develop a phenomenological theory describing this type of skyrmion.

\end{abstract}

%\pacs{PACS numbers here}
% PACS, the Physics and Astronomy Classification Scheme.

%\keywords{Keywords here}

\maketitle

%%%%%%%%%%%%%%%%%%%%%%%%%%%%%%%%%%%%%%%%%%%%%

\section{Introduction}
\label{Introduction}

In magnetic compounds lacking inversion symmetry, the handedness of the underlying crystal structure induces an asymmetric exchange coupling called the \textit{Dzyaloshinskii-Moriya} (DM) interaction~\cite{Dz64} which stabilizes long-period spatial modulations of the magnetization with a fixed rotation sense~\cite{Dz64, Bak80}. There has been a renewed interest in  chiral helimagnetism in the last few years since the discovery of two-dimensional localized modulations called \textit{chiral skyrmions}~\cite{Bogdanov89,Bogdanov94, Romming13, Leonov16}.

In most nonlinear physical systems, similar two-dimensional localized states are radially unstable and collapse spontaneously into linear singularities.  Chiral interactions provide a unique stabilization
mechanism, protecting two-dimensional localized states from this instability \cite{Bogdanov94,Leonov16}. That is why non-centrosymmetric magnets and other chiral condensed matter systems are of special interest in fundamental physics and mathematics as a particular class of materials where skyrmions can exist~\cite{Rossler06, Melcher15}. Chiral magnetic skyrmions are also considered promising objects for various spintronic applications, notably racetrack computer memories~\cite{Kiselev11,Fert13, Parkin15}.

\begin{figure} 
 \includegraphics[width=0.95\columnwidth]{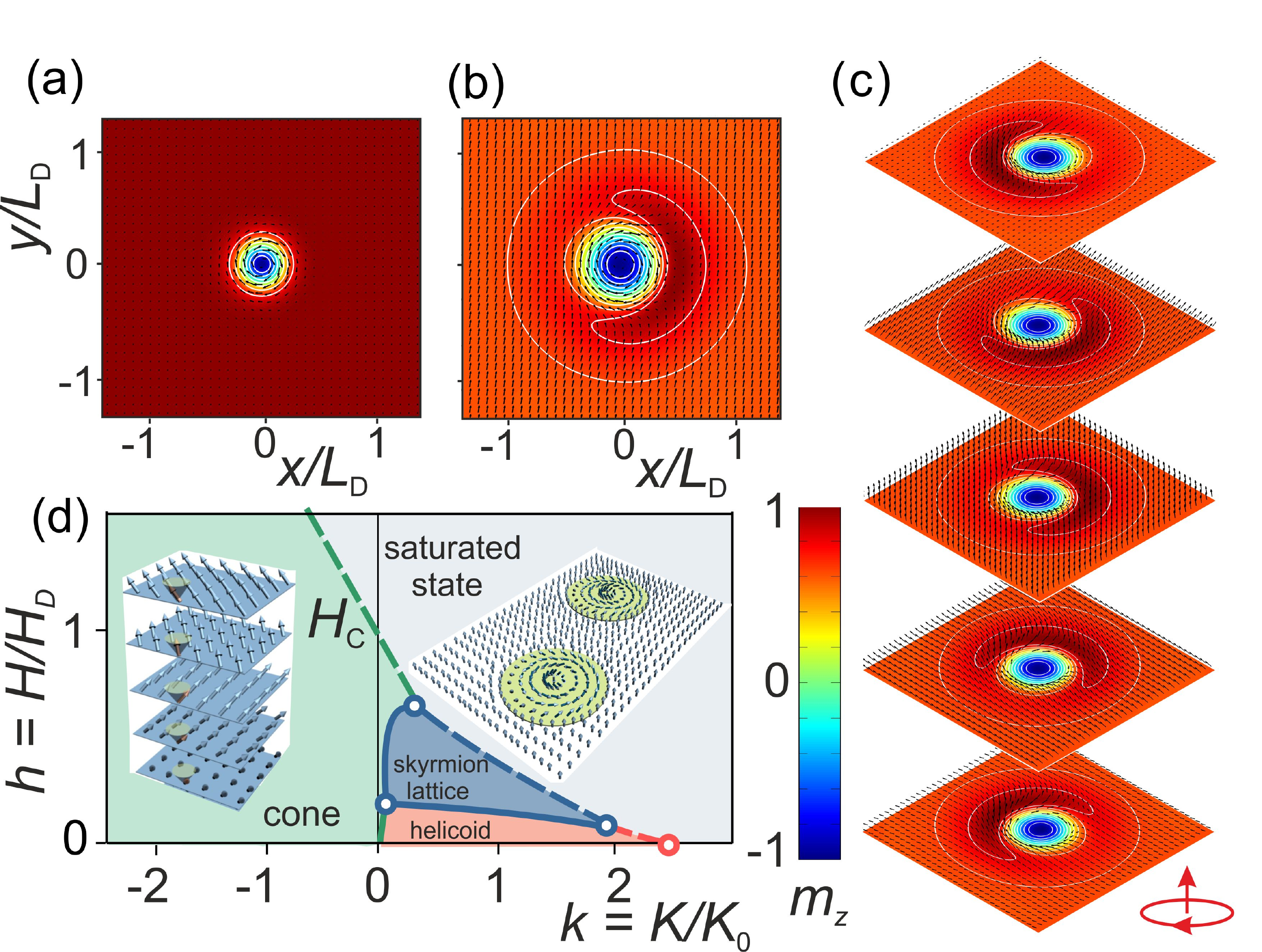}
\caption{\label{crankshaft} Magnetic structures of chiral skyrmions with zero anisotropy. (a--c) show color plots of the out-of-plane magnetic moment, $m_z (x, y)$, for (a) an axisymmetric skyrmion in the saturated state ($h$ = 1.1) and (b) a non-axisymmetric skyrmion in the cone phase ($h$ = 0.6). (c) shows a stack of magnetization planes $m_z(x,y)$ with different values of $z$ for the non-axisymmetric skyrmion. In the generic phase diagram of non-centrosymmetric ferromagnets (d), axisymmetric skyrmions (a) exist within the saturated phase and transform continuously into non-axisymmetric skyrmions (b) during the transition into the cone phase at the critical line $H_{C}$.}
\end{figure}

The generic phase diagram for the magnetic states which can occur in in non-centrosymmetric magnets is shown in Fig.~\ref{crankshaft}(d). Skyrmions occur as lattices or as isolated particles within a different magnetic phase. The chiral skyrmions that were theoretically introduced in Refs.~\onlinecite{Bogdanov89, Bogdanov94, Bogdanov95} and experimentally investigated in nanolayers of chiral ferromagnets \cite{Romming13,Leonov16} were embedded in the magnetically saturated phase in which all the atomic spins are parallel. Recently, fundamentally different solitonic states embedded in the cone phase of chiral ferromagnets have been investigated mathematically \cite{LeonovJPCM16}. Unlike the axisymmetric skyrmions, these three-dimensional chiral solitons are inhomogeneous along their axes and asymmetric in the basal planes as shown in Fig. 1(a-c)\cite{LeonovJPCM16}. Whereas skyrmions embedded in the saturated phase repel one another, these skyrmions are mutually attractive and so tend to produce clusters of skyrmions~\cite{LeonovJPCM16, LeonovAPL16}. 

Non-axisymmetric skyrmions have been investigated theoretically in bulk~\cite{LeonovJPCM16} and confined chiral helimagnets~\cite{LeonovAPL16} but no experimental observations of these states have been reported to date. Here we show the first observation of attractive skyrmions in the cone phase of the cubic helimagnet Cu$_2$OSeO$_3$ by presenting transmission electron micrographs of skyrmion clusters and provide a theoretical description of these results.

%%%%%%%%%%%%%%%%%%%%%%%%%%%%%%%%%%%%%%%%%%%%%

\section{Theory}
\label{Theory}

Chiral solitons and modulated phases are described mathematically by equations which minimize the energy functional for a chiral ferromagnet~\cite{Dz64,Leonov16}:

\begin{eqnarray}
w = A (\mathbf{grad}\,\mathbf{M})^2 + 
K (\mathbf{M}\cdot \mathbf{n})^2
- \mu_0  \mathbf{M}\cdot\mathbf{H} + w_D, \,\,  
\label{density}
\end{eqnarray}
where $\mathbf{M}$ is the magnetization $\mathbf{M}=M(\sin\theta\cos\psi;\sin\theta\sin\psi;\cos\theta)$ with $\theta$ being the polar angle and $\psi$ being the azimuthal angle between each magnetic moment and the applied magnetic field $\mathbf{H}$ which points in the $z$ direction. $A$ is the exchange stiffness constant, $K$ is the uniaxial anisotropy constant and $\mathbf{n}$ is the unity vector along uniaxial anisotropy axis. The DM energy functionals $w_D$ are composed of Lifshitz invariants $\mathcal{L}^{(k)}_{i,j} = M_i \partial M_j/\partial x_k - M_j \partial M_i/\partial x_k $~\cite{Dz64,Bogdanov89} where $x$ is the spatial coordinate.

Denoting the distance from the skyrmion axis as $\rho$, skyrmions in the cone phase approach the solutions for the cone phase~\cite{Butenko10,Wilson14} ($\theta_c, \psi_c$) at large distances from the axis \cite{Leonov16}:

\begin{eqnarray}
\theta_{\rho\rightarrow \infty} = \theta_c = \arccos \left(H/H_C \right), \, 
\psi_{ \infty} = \psi_c =   2\pi z/L_D, \, \, 
\label{cone}
\end{eqnarray}
where $H_C = H_D \left( 1- K/K_0 \right)$ is the magnetic field above which the saturated state forms and $H_D =D^2 M/(2A)$. The pitch of the cone phase is $L_D = 4\pi A/|D|$ and $K_0 = D^2 /(4A)$ where $D$ is the Dzyaloshinskii-Moriya (DM) coupling energy.

For the important case of non-centrosymmetric cubic ferromagnets (with space group P2$_1$3), the DM energy functional is reduced to the isotropic form~\cite{Bak80}: $w_D = D(\mathcal{L}_{yx}^{(z)}+\mathcal{L}_{xz}^{(y)}+\mathcal{L}_{zy}^{(x)})=D\,\mathbf{M}\cdot \mathrm{rot}\mathbf{M}$.  The calculated phase diagram for this case is representative for a whole
class of non-centrosymmetric ferromagnets and shown in Fig.~\ref{crankshaft}(d)~\cite{Butenko10}.  In this phase diagram, axisymmetric skyrmions exist as ensembles of weakly repulsive strings within the magnetically saturated phase and non-axisymmetric skyrmions inhabit the cone phase. During the phase transition from the saturated to the cone phase through the critical line $H_C$ (Fig. \ref{crankshaft} (d)), axisymmetric skyrmions continuously transform into non-axisymmetric skyrmions. The structure of these non-axisymmetric skyrmions is discussed in detail in Supplemental Information I~\cite{supp} and unlike the axisymmetric skyrmions, the point at which $\theta=\pi$ is no longer lies on the central axis of the skyrmion but its position oscillates on moving in $z$ from one layer to the next. Experimentally, we produced individual non-axisymmetric skyrmions and their clusters by starting in the skyrmion lattice phase and increasing the magnetic field so the lattice fragments during the first-order transition to the cone phase. 

%%%%%%%%%%%%%%%%%%%%%%%%%%%%%%%%%%%%%%%%%%%%%

\section{Experimental Methods}
\label{methods}

Copper oxy-selenite (Cu$_2$OSeO$_3$) is a cubic helimagnet with a characteristic helical period~\cite{Seki12} of 63~nm. {\it Ab-initio} calculations show that this helimagnetism is induced by the DM interaction~\cite{Yang12}. To acquire images of skyrmions, a single crystal was thinned to electron transparency (about 70~nm) by the conventional process of mechanical grinding and argon-ion polishing on the (110) face. Lorentz transmission electron microscopy (LTEM) was conducted using an FEI Tecnai F20 electron microscope and the sample cooled {\it in-situ} using a liquid-helium cooled holder with a base temperature of 10~K. Skyrmions appear as black or white circles in the images produced using this technique. The images also show white linear features which are not magnetic but are parts of the specimen surface which had peeled off and rolled into tiny tubes. We have not encountered this with the preparation of other materials. (See Supplemental Information II for full experimental methods~\cite{supp}.)
 
\section{Experimental Results}

Fig.~\ref{clusters2} shows frames from one of several videos we acquired (see Supplemental Information III~\cite{supp}) showing two skyrmion clusters (outlined in red) embedded in a host phase that produced no contrast in the image. At lower fields, the skyrmion lattice filled the whole sample. The smaller cluster contained 7 skyrmions and the larger, 13. There does not appear to be a preferred number of skyrmions in a cluster and in other videos we observed a single skyrmion as well as clusters with 6, 18 and 21 skyrmions. The fact that the skyrmions form clusters demonstrates that the interaction between them must be attractive and later we show theoretically that skyrmions embedded in the cone phase can be attractive whereas skyrmions embedded in the saturated state are purely repulsive.

\begin{figure} 
\includegraphics[width=0.95\columnwidth]{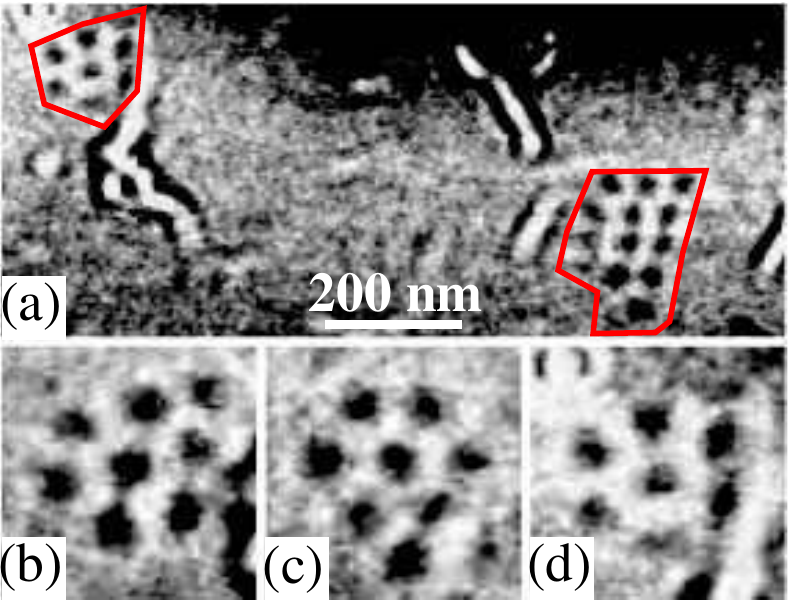}
\caption{\label{clusters2} (a) LTEM images showing two skyrmion clusters (outlined in red) embedded in the cone phase of Cu$_2$OSeO$_3$ at 11~K in an out-of-plane applied magnetic field of $\mu_0 H$ = 116~mT. (b)--(d) show changes in the shape of the left-hand cluster with (c) taken 0.5~s and (d) taken 34.4~s after (b)
}
\end{figure}

\begin{figure}  
\includegraphics[width=0.95\columnwidth]{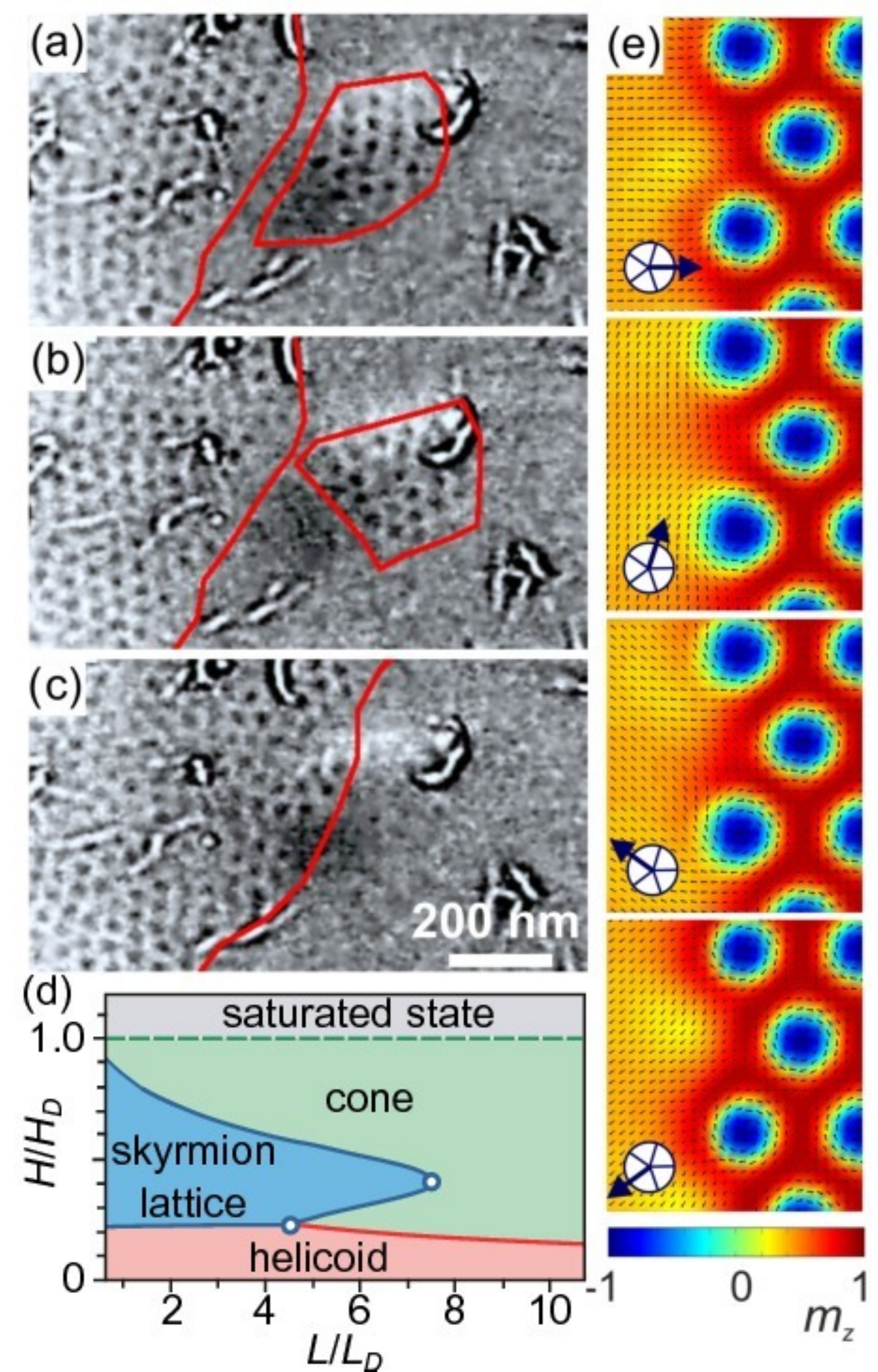}
\caption{\label{clusters1} LTEM images of a Cu$_2$OSeO$_3$ nanolayer of thickness $L$ = 70~nm at 12~K in an out-of-plane applied magnetic field of $\mu_0 H$ = 116~mT. (a--c) show the skyrmion lattice phase on the left coexisting with the cone phase and the phase boundary delineated in red. A skyrmion cluster, outlined in red, merges with the skyrmion lattice phase over time with (b) taken 1.6~s after (a) and (c) 21.2~s after (a) during the first-order transition between the cone and skyrmion lattice phases. (d) A sketch of the phase diagram of cubic helimagnet nanolayers (adapted from Ref.~\onlinecite{LeonovPRL16}) includes areas with the helicoid, cone, and skyrmion lattices phases separated by first-order transition lines (solid lines) and the saturated state separated from the cone phase by a second-order transition line (dashed). (e) The calculated domain wall between the coexisting skyrmion lattice and cone phases at the first-order transition line.}
\end{figure}

Figs.~\ref{clusters1}(a)--(c) were taken under the same conditions from a different region of the sample and show a cluster of 30 skyrmions moving across the host phase and merging with the skyrmion lattice phase over 21~s. The boundaries of the cluster and the edge of the skyrmion lattice phase are delineated by red lines and by comparing panels (b) and (c), it can be seen that the phase boundary advances after merging and that the skyrmions in the cluster have spread evenly across the boundary.

The host phase appears featureless in the images so it is either the cone or saturated state since neither produce any contrast variation. Whilst the image alone does not allow us to identify the host phase, it is almost certainly the cone phase. It is clear from Figs.~\ref{clusters1}(a)--(c) that the experimental conditions are close to the phase boundary of the skyrmion lattice. Both the phase diagram for bulk Cu$_2$OSeO$_3$ derived by neutron diffraction~\cite{Adams12} and the theoretical phase diagram for thin cubic helimagnets in Fig.~\ref{clusters1}(d)~\cite{LeonovPRL16} indicate that if the material is initially in the skyrmion lattice phase, an increase in magnetic field will first create the cone phase and any coexistence should be between these two phases. Furthermore, coexisting phases like those we see here can only result from a first order transition whereas a transition from the skyrmion lattice to the saturated state should be a second order process which occurs via the gradual expansion of the period of the skyrmion lattice~\cite{Bogdanov94, LeonovPRL16}.

In all of the videos we recorded, the skyrmions were in constant motion as shown in Fig.~\ref{clusters2}(b)--(d). Similar motion has been reported by Mochizuki {\it et al.}~\cite{Mochizuki14} who attribute this to the heating of the specimen by the electron beam. We suggest instead that it may be caused by the specimen charging under the electron beam as coating the sample in a thin layer of carbon to improve its electrical conductivity slowed the movement of the skyrmions. Such charging can occur if the sample is insulating, like Cu$_2$OSeO$_3$, or not well grounded. Mochizuki {\it et al.} calculated that a steady flow of electrons through the sample from the electron beam was three orders of magnitude too low to move skyrmions via the spin-torque effect but it is possible that bursts of current caused by the specimen charging and discharging may be sufficient to cause this movement.

Areas with the cone and skyrmion lattice phases are separated by domain walls. The calculated contour plots of such a domain wall is presented in Fig.~\ref{clusters1}(e). The frontal parts of skyrmion
cores in the wall have a similar structure as those in non-axisymmetric skyrmions and play the role of nuclei for individual non-axisymmetric skyrmions. The attractive interaction between such skyrmions~\cite{LeonovJPCM16} explains the formation of skyrmion clusters observed in our experiments.

%%%%%%%%%%%%%%%%%%%%%%%%%%%%%%%%%%%%%%%%%%%%%%%%%%%

\section{Discussion}
\label{Discussion}

A non-axisymmetric skyrmion can be thought as the result of a continuous transformation of an axisymmetric skyrmion during the phase transition from the saturated to the cone phase (Fig. 1(d)). The  equilibrium structure of a non-axisymmetric skyrmion is reached by the compatibility of the axisymmetric skyrmion core with the transversely modulated cone phase (Fig. \ref{crankshaft}(a-c)) ~\cite{LeonovJPCM16,LeonovAPL16}. The skyrmion core is separated from the host phase
by a broad asymmetric ``shell''.

\begin{figure}  
 \includegraphics[width=0.95\columnwidth]{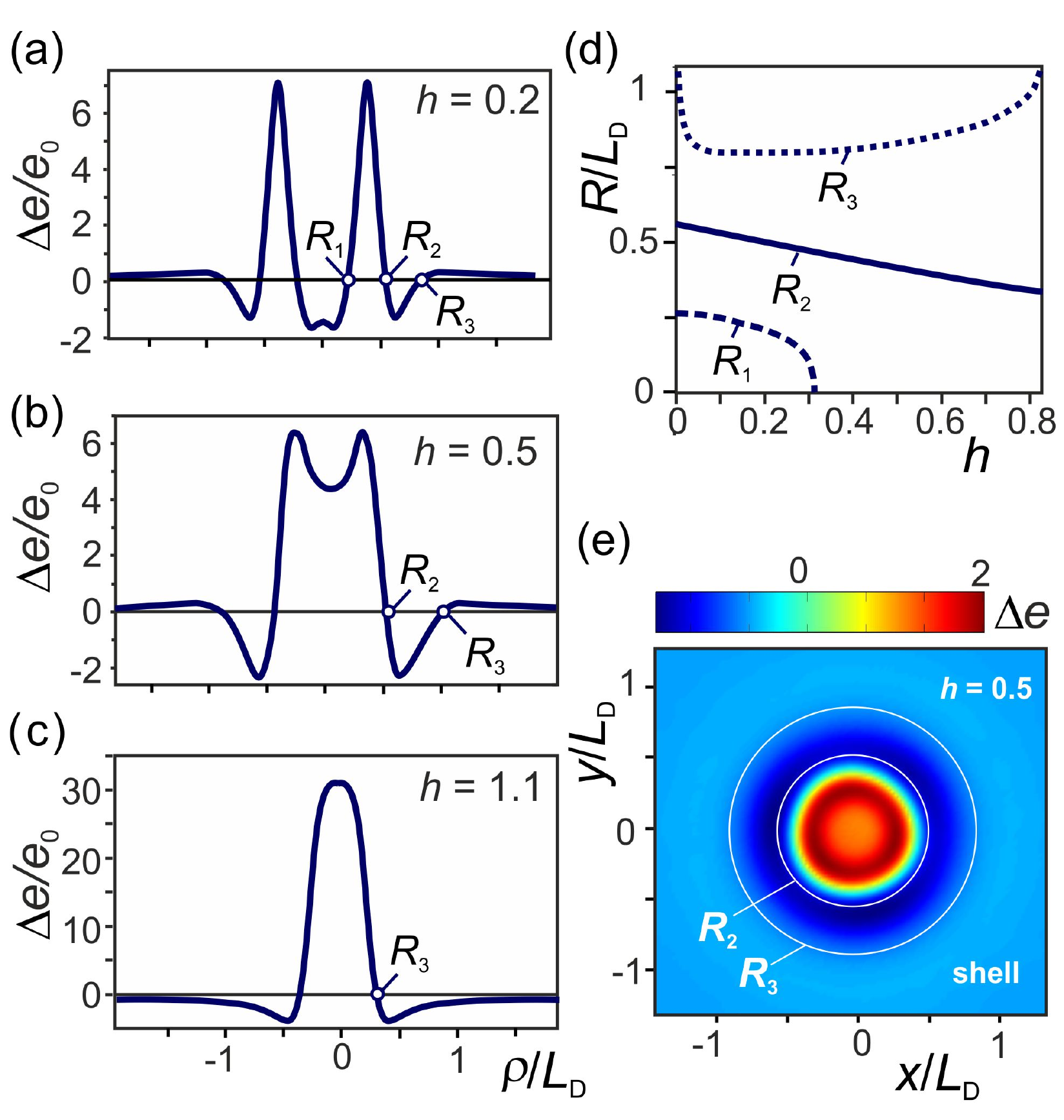}
\caption{(color online). (a--c) Radial energy density profiles $\Delta e(\rho)$ (see Eqn. \ref{energy}) at zero anisotropy with different values of the applied field plotted in units $e_0 = 4\pi D $. (d) Characteristic radii $R_i$ are plotted as functions of the applied field. In the cone phase (a,b) the skyrmion core is enclosed by positive ``shells'' ($\rho > R_3$) providing the attractive inter-skyrmion potential.  Negative asymptotics of the radial energy density for axisymmetric skyrmions ($h > 1$) impose the repulsive skyrmion-skyrmion interaction (c). (e) The contour plot $e(x, y)$ calculated for $h = 0.5$ shows the central part with a large positive energy ($\rho < R_2$), the negative energy belt ($R_2 < \rho < R_3$), and the extended shell ($\rho > R_2$).
\label{crank4}
}
\end{figure}

The numerical calculations presented in Fig. \ref{crank4} elucidate the main features of non-axisymmetric skyrmions and their bound states. The calculated radial skyrmion energy densities 

\begin{eqnarray}
e(\rho) = (2 \pi L_D)^{-1} 
\int_0^{L_D} dz \int_0^{2\pi} d \varphi w_s (\theta, \psi) 
\label{energy}
\end{eqnarray}
are plotted as functions $\Delta e(\rho) = (e (\rho) - e_{\mathrm{cone}})$ for different values of the applied field (Fig.~\ref{crank4}) where $w_s(\theta, \psi)$ is the energy density (Eqn.~\ref{density}) for an isolated non-axisymmetric skyrmion and $e_{\mathrm{cone}}$ is energy density (Eqn.~\ref{energy}) calculated for the cone phase (Eqn. \ref{cone}). It should be noted that as the host cone phase and embedded non-axisymmetric skyrmions are modulated along the film thickness $L$, the skyrmion energy density (Eqn.~\ref{energy} and Fig.~\ref{crank4}(e)) and the inter-skyrmion coupling depend on the confinement ratio, $\nu = L/L_D$.  These subtle effects could be a subject of future theoretical and experimental investigations.

The characteristic lengths $R_1$, $R_2$, $R_3$ indicate several distinct regions in the radial energy density profiles $\Delta e(\rho)$ (Fig.~{\ref{crank4}}(c)--(e)).  The functions $R_i (h)$ are plotted in Fig.~\ref{crank4}(d). For axisymmetric skyrmions (Fig.~\ref{crank4}(c)), the energy densities $\Delta e(\rho)$ consist of a positive energy ``bag'' located in the skyrmion center ($\rho < R_2$) surrounded by extended areas with negative energy density, where the DM coupling dominates \cite{Bogdanov94}. Negative asymptotics of the radial energy densities ($\Delta e(\rho) < 0$ for $\rho >> 1$ predetermines the \textit{repulsive} inter-soliton potential for axisymmetric skyrmions \cite{Bogdanov94}. For non-axisymmetric skyrmions (Fig.~\ref{crank4}(a) and (b)) the energy densities $\Delta e(\rho)$ are
positive at large distances from the skyrmion center ($\rho > R_3$). These areas correspond to ``shells'' separating the skyrmion core from the cone phase. The positive energy density of the shell
leads to attractive interactions between non-axisymmetric skyrmions \cite{LeonovJPCM16}. In other words, the attractive interaction between skyrmions in the cone phase is explained by the excessive energy density of the asymmetric shell compared with the skyrmion core and the cone phase~\cite{LeonovJPCM16}. Importantly, the equilibrium distribution of the magnetization in the shell determines the material parameters of the cluster: the bound energy and the distance between the constituent skyrmions.

Another type of attractive chiral skyrmion has been investigated theoretically in the precursor region of cubic helimagnets~\cite{Wilhelm11,Wilhelm12}. It was shown that due to the
``softening'' of the magnetization near the ordering temperature, the skyrmion-skyrmion coupling acquires an oscillatory character and promotes the formation of skyrmion clusters in this region.

The solutions for two-dimensional ``baby'' skyrmions with an oscillatory inter-particle potential have also been derived within the canonical Skyrme model \cite{LeonovNcomm15}. In magnetism the Skyrme
model is applied to describe a group of magnetic compounds with competing exchange interactions. {\it Ab-initio} calculations of attractive two-dimensional localized states in these magnetic systems
have been carried out by Rozsa {\it et al.} \cite{Rozsa16}. It was also found that the solutions for attractive baby skyrmions exist in modified Faddeev-Skyrme models with a Lennard-Jones type potential
term describing a short-range repulsion and a long-range attraction~\cite{Salmi15}.

The oscillatory vortex-vortex interaction attributed to type-II superconductors with small values of the Ginzburg-Landau parameter~\cite{Hubert71} leads to the first-order transition from the superconducting state into the Abrikosov vortex phase accompanied with the multidomain states of the competing phases~\cite{Essmann71}. Vortex clusters stabilized by the attractive inter-vortex coupling
have been also observed in MgB$_2$ and Sr$_2$RuO$_4$ \cite{Moshchalkov09, Gutierrez12, Curran11, Garaud14}. The attractive skyrmions in the cone phase of non-centrosymmetric ferromagnets represent an alternative to solitons with the oscillatory inter-particle potential investigated in Refs.~\onlinecite{Wilhelm11, Wilhelm12, LeonovNcomm15, Rozsa16, Hubert71, Essmann71}.

In conclusion, we report the first direct observations of clusters of attractive skyrmions embedded in the cone phase of a non-centrosymmetric ferromagnet. The clusters were generated by the magnetic-field-induced fragmentation of the skyrmion lattice during the first-order transition to the cone phase. This investigation used Cu$_2$OSeO$_3$ but the same method could be used to investigate skyrmion clusters in the cone phases of other non-centrosymmetric ferromagnets.

\acknowledgements
The authors are grateful to  E. Babaev, K. Inoue, D. McGrouther, T. Monchesky and Y. Togawa for useful discussions. This work was funded by the Royal Society (United Kingdom) and the United Kingdom Engineering and Physical Sciences Research Council (EPSRC), grant number EP/N032128/1. M.C.H. and G.B. also acknowledge financial support from the EPSRC grant number EP/M028771/1. A.O.L. acknowledges the Japan Society for the Promotion of Science (JSPS) Core-to-Core Program, Advanced Research Networks and the JSPS Grant-in-Aid for Research Activity Start-up 17H06889. A.O.L. thanks Ulrike Nitzsche for technical assistance.

%---------- Bibliography ----------

%\bibliography{sky}% Produces the bibliography via BibTeX.

\providecommand{\noopsort}[1]{}\providecommand{\singleletter}[1]{#1}%

\end{document}